\documentclass[12pt]{article}
\usepackage{graphicx}
\usepackage{amsmath,amssymb}
\usepackage{bm}
\usepackage{graphicx, color}
\usepackage{wrapfig}
\begin{document}
\title{
\begin{flushright}
\ \\*[-80pt] 
\begin{minipage}{0.2\linewidth}
\normalsize
\end{minipage}
\end{flushright}
{\Large \bf   Squark flavor mixing and CP asymmetry \\
 of neutral $B$ mesons at LHCb
\\*[20pt]}}

\author{
\centerline{Atsushi~Hayakawa$^{1,}$\footnote{E-mail address: hayakawa@muse.sc.niigata-u.ac.jp}, \ \
Yusuke~Shimizu$^{2,}$\footnote{E-mail address: shimizu@muse.sc.niigata-u.ac.jp}} \\ 
\centerline{Morimitsu~Tanimoto$^{2,}$\footnote{E-mail address: tanimoto@muse.sc.niigata-u.ac.jp}, \ \
Kei~Yamamoto$^{1,}$\footnote{E-mail address: yamamoto@muse.sc.niigata-u.ac.jp}}
\\*[20pt]
\centerline{
\begin{minipage}{\linewidth}
\begin{center}
$^1${\it \normalsize
Graduate~School~of~Science~and~Technology,~Niigata~University, \\ 
Niigata~950-2181,~Japan }
\\*[4pt]
$^2${\it \normalsize
Department of Physics, Niigata University,~Niigata 950-2181, Japan }
\end{center}
\end{minipage}}
\\*[70pt]}

\date{
\centerline{\small \bf Abstract}
\begin{minipage}{0.9\linewidth}
\vskip  1 cm
\small
 The CP violation of the neutral $B$ meson is the important phenomenon
to search for the new physics. 
The like-sign dimuon charge asymmetry observed  by  the D$\O$ 
Collaboration indicates the CP-violating new physics 
in the $B_s-\bar B_s$  mixing.
On the other hand,  LHCb  observed 
 the CP-violating asymmetry in $B_s^0\to J/\psi \phi$ 
and  $B_s^0\to J/\psi f_0(980) $, which 
 is consistent with the SM prediction.
However, there is still room for new physics 
of  the CP violation.
The CKMfitter has presented the allowed region of the
 new physics parameters  taking  account of the LHCb data.
 Based on these results, we  discuss  the effect of the squark flavor mixing 
on the CP violation in the $B_d$ and $B_s$ mesons.
We predict  asymmetries in the non-leptonic decays 
 $B_d^0\to \phi K_S$, $B_d^0\to \eta 'K^0$, 
$B_s^0\to\phi \phi$ and $B_s^0\to\phi\eta '$.
\end{minipage}
}

\begin{titlepage}
\maketitle
\thispagestyle{empty}
\end{titlepage}

\section{Introduction}
\label{sec:Intro}

The CP violation in the $K$ and $B_d$ mesons has been successfully
understood 
within the framework of the standard model (SM),
 so called Kobayashi-Maskawa (KM) model \cite{Kobayashi:1973fv}.
The source of the CP violation is the KM phase 
in the quark sector with three families. 
Until now, the KM phase has successfully described 
the experimental data  of the CP violation of $K$ and $B_d$ mesons.

 However, there could be new sources of the CP violation if the SM is 
extended to the supersymmetric (SUSY) models. The CP-violating phases appear
in  soft scalar mass matrices. These phases  contribute to flavor changing 
neutral  currents  with the CP violation.
Therefore, we should examine carefully CP-violating phenomena
in the neutral mesons.

The Tevatron experiments have searched signals of the CP violation
in the $B$ mesons. 
 Recently, the D$\O$  Collaboration reported 
the interesting result of the like-sign dimuon charge asymmetry
$A_{sl}^b({\rm D}\O)=-(7.87\pm 1.72\pm 0.93)\times 10^{-3}$
 \cite{Abazov:2010hj}. 
This result is larger than the SM prediction 
$A_{sl}^b({\rm SM})=(-2.3 ^{+0.5}_{-0.6}) \times 10^{-4}$ \cite{Abazov:2010hj,Lenz:2006hd}
 at the $3.9~\sigma$ level,
which indicates the CP-violating 
  new physics in the $B_s$-$\bar B_s$  mixing
\cite{Ligeti:2006pm,Ligeti:2010ia}.  

On the other hand,
the LHCb~\cite{LHCb:2011ab,LHCb:2011aa} and the CDF \cite{CDF:2011af} 
observed the CP-violating phase $\phi _s$ in the non-leptonic
 decays of $B_s^0\to J/\psi \phi $ 
and  $B_s^0\to J/\psi f_0(980) $.
Those results are  consistent with 
the SM prediction.  However, there is still room 
for new physics on  the CP violation of the $B$ meson.
Actually, the CKMfitter has presented the allowed region of the
 new physics parameters  taking into account of LHCb data 
\cite{CKMfitter,Ligeti}. (See also the work in Ref.~\cite{Altmannshofer:2011iv}.) 

The typical new physics  is 
the  gluino-squark mediated flavor changing process
 based on the SUSY model
 \cite{King:2010np}-\cite{Ishimori:2011nv}.
Relevant mass insertion parameters can 
explain the anomalous CP violation in the $B_s$ meson. 
 In this paper, we  discuss  the effect of the squark flavor mixing 
on the CP violation in the non-leptonic decays
of $B_d$ and $B_s$
taking  account of the recent LHCb experimental data.
Then, the CP-violating phases of  the squark flavor mixing are constrained 
by the chromo electric dipole moment (cEDM) of strange quark 
\cite{Hisano:2003iw,Hisano:2004tf,Hisano:2008hn}.
The prediction of 
 asymmetries in  the penguin dominated decays is the crucial test
 of  the squark flavor mixing.  We predict the asymmetries of 
 $B_d^0\to \phi K_S$, $B_d^0\to \eta 'K^0$, 
$B_s^0\to\phi \phi$ and $B_s^0\to\phi\eta '$ decays.

In section 2,
we summarize the recent  experimental  situation  
 in  the CP violation of the neutral $B$ mesons. 
In section 3,
we discuss  the contribution of the squark favor mixing on 
the $B$ mesons.
  We also discuss the constraints from the $b\to s\gamma$ process
 and the cEDM of the strange quark.
 In section 4, we present the numerical result of 
the CP violation in the non-leptonic decays  of $B$ mesons.
Section 5 is devoted to the summary and discussion.

\section{New physics of  CP violation in  $B_q$-$\bar B_q$ system}
\label{sec:Deviation}
Let us discuss the possible contribution of the new physics
on the $B_q$-$\bar B_q (q=d,s)$ system.
The Tevatron experiment reported about the CP violation 
in like-sign dimuon charge asymmetry $A_{sl}^b$,
which  is defined as~\cite{Abazov:2010hj,Grossman:2006ce}
\begin{equation}
A_{sl}^b\equiv \frac{N_b^{++}-N_b^{--}}{N_b^{++}+N_b^{--}}
=(0.506\pm 0.043)a_{sl}^d+(0.494\pm 0.043)a_{sl}^s~.
\end{equation}
Here,  $N_b^{\pm \pm }$ is the number of events of 
$b\bar b\to \mu ^{\pm \pm }X$, 
and the "wrong-sign" charge asymmetry $a_{sl}^q$ of $B_q\to \mu ^-X$ decay 
is defined as 
\begin{equation}
a_{sl}^q\equiv \frac{\Gamma (\bar B_q^0\rightarrow \mu ^+X)-\Gamma (B_q^0\rightarrow \mu ^-X)}
{\Gamma (\bar B_q^0\rightarrow \mu ^+X)+\Gamma (B_q^0\rightarrow \mu ^-X)}
\simeq \text{Im} \left (\frac{\Gamma _{12}^q}{M_{12}^q}\right ),
\end{equation}
where
$M_{12}^q$ and $\Gamma _{12}^q$ are dispersive and 
absorptive part in the effective Hamiltonian of 
the $B_q$-$\bar B_q$ system, respectively.
The SM prediction of $A_{sl}^b$ is given as~\cite{Abazov:2010hj}
\begin{equation}
A_{sl}^b(\text{SM})=(-2.3_{-0.6}^{+0.5})\times 10^{-4}, 
\label{AslbSM}
\end{equation}
which is calculated from~\cite{Lenz:2006hd}
\begin{equation}
a_{sl}^d(\text{SM})=(-4.8_{-1.2}^{+1.0})\times 10^{-4},
\qquad a_{sl}^s(\text{SM})=(2.06\pm 0.57)\times 10^{-5}.
\end{equation}
The D$\O $ Collaboration reported $A_{sl}^b$ with 9.0 fb$^{-1}$ data set 
as~\cite{Abazov:2010hj}
\begin{equation}
A_{sl}^b(\text{D\O })=-(7.87\pm 1.72\pm 0.93)\times 10^{-3}, 
\end{equation}
which shows $3.9~\sigma $ deviation from the SM prediction of Eq.~(\ref{AslbSM}). 

Therefore, we consider the new physics  beyond the SM. 
The contribution of new physics to the dispersive part $M_{12}^q$ 
is parameterized as 
\begin{equation}
M_{12}^q=M_{12}^{q,\text{SM}}+M_{12}^{q,\text{NP}}=
M_{12}^{q,\text{SM}}(1+h_qe^{2i\sigma _q})~, \quad (q=d,s)
\end{equation}
where $M_{12}^{q,\text{NP}}$ are new physics contribution, 
and the SM contribution $M_{12}^{q,\text{SM}}$ are given as \cite{sanda}
\begin{equation}
M_{12}^{q,\text{SM}}=\frac{G_F^2M_{B_q}}{12\pi ^2}M_W^2(V_{tb}V_{tq}^*)^2\hat \eta _BS_0(x_t)f_{B_q}^2B_q~.
\end{equation} 
The SM contribution to the absorptive part $\Gamma _{12}^q$ is dominated 
by tree-level decay $b\to c\bar c s$, $\tau ^+\tau ^-s$, and etc. 
Then, we assume $\Gamma _{12}^q=\Gamma _{12}^{q,\text{SM}}$.
 Numerical values  of the new physics parameters  $h_q$ and $\sigma _q$
 have been obtained by the CKMfitter \cite{CKMfitter,Ligeti}.


Let us discuss the effect of the new physics 
 in the non-leptonic decays of $B$ mesons. 
The time dependent  CP asymmetry decaying into the final state $f$, 
 which is defined as \cite{Aushev:2010bq}
\begin{equation}
\mathcal{S}_{f}=\frac{2\text{Im}\lambda _{f}}{|\lambda_{f}|^2+1}\ ,
\label{sf}
\end{equation}
where 
\begin{equation}
\lambda_{f}=\frac{q}{p} \bar \rho\ , \qquad 
\bar \rho \equiv
\frac{\bar A(\bar B_q^0\to f)}{A(B_q^0\to f)}.
\end{equation}
In the decay of  $B_d^0\to J/\psi  K_S$, we take
\begin{equation}
\lambda_{J/\psi  K_S}=
-e^{-i\phi _d},\quad \phi _d
=2\beta_d+\text{arg}(1+h_de^{2i\sigma _d}),
\end{equation}
by putting  $|\bar \rho |=1$ and $q/p\simeq \sqrt{M_{12}^{q*}/M_{12}^q}$,
where the phase $\beta_d $ is given  in the SM.
The CKMfitter provided the allowed region of $h_d$ and $\sigma_d$,
where the central value is \cite{CKMfitter,Ligeti}
\begin{equation}
h_d\simeq 0.3,\qquad \sigma _d\simeq 1.8 \ {\rm rad}.
 \label{hdsigmaLHCb}
\end{equation}
 Since  penguin processes are dominant  in  the case of
  $f=\phi K_S, \eta ' K^0$, the loop induced new physics
 could  contribute considerably 
on the   CP violation of  those  decays.
 Then, those  $\mathcal{S}_f$ is not any more same 
as $\mathcal{S}_{J/\psi K_S}$
 due to $|\bar \rho |\not = 1$.
 Those predictions provide us good tests for the new physics.

In the decay of  $B_s^0\to J/\psi\phi$, we have 
\begin{equation}
\lambda _{J/\psi \phi }=
e^{-i\phi _s},\qquad \phi _s
=-2\beta_s+\text{arg}(1+h_se^{2i\sigma _s}),
\end{equation}
where $\beta _s$ is given  in the SM.

Recently the  LHCb~\cite{LHCb:2011ab} presented the observed 
 CP-violating phase $\phi _s$ in $B_s^0\to J/\psi \phi $ 
and  $B_s^0\to J/\psi f_0(980) $ decays using about 340 pb$^{-1}$ of data. 
The combination of these  results lead to
\begin{equation}
\phi _s=0.07\pm 0.17\pm 0.06~ \text{rad}.
\label{phis}
\end{equation}
On the other hand, the SM prediction is  \cite{CKMfitter}
\begin{equation}
\phi _s^{J/\psi \phi ,SM}=-2\beta_s=-0.0363\pm 0.0017~\text{rad}.
\label{SMBs}
\end{equation}


Taking  account of these data, the CKMfitter has presented
the allowed values of $h_s$ and $\sigma _s$ ~\cite{CKMfitter,Ligeti}.
The allowed region is rather large including zero values.
In order to investigate possible contribution of the new physics,  we take 
the central values
\begin{equation}
h_s= 0.1,\qquad \sigma _s=0.9 - 2.2 \ \text{rad},
\label{hsigmaLHCb}
\end{equation}
as a typical parameter set in our work.

\section{Squark flavor mixing}
\label{sec:Squark}
As the new physics contributing on  the CP violation of the neutral $B$ meson,
 we study  the effect  of the squark flavor mixing in SUSY.
Let us consider the flavor structure of squarks, which gives
 the flavor changing neutral currents.
When three families correspond to a triplet of 
a certain   flavor symmetry, for example $A_4$ and $S_4$ 
\cite{Ishimori:2010au},
the  squark mass matrix is diagonal with three degenerate masses
 in the supersymmetric limit.
Then, the SUSY breaking induces soft SUSY
breaking terms such as squark  masses and scalar trilinear couplings, 
i.e. the so-called A-terms.
The breaking of the flavor symmetry
 gives the small soft masses compared with the diagonal ones
 in the squark mass matrices.
Therefore, in the super-CKM basis, we parametrize 
 the soft scalar masses squared 
$M^2_{\tilde d_{LL}}$, $M^2_{\tilde d_{RR}}$, 
$M^2_{\tilde d_{LR}}$, and $M^2_{\tilde d_{RL}}$ for the down-type squarks
as follows:
\begin{align}
M^2_{\tilde d_{LL}}&=m_{\tilde q}^2
\begin{pmatrix}
1+(\delta _d^{LL})_{11} & (\delta _d^{LL})_{12} & (\delta _d^{LL})_{13} \\
(\delta _d^{LL})_{12}^* & 1+(\delta _d^{LL})_{22} & (\delta _d^{LL})_{23} \\
(\delta _d^{LL})_{13}^* & (\delta _d^{LL})_{23}^* & 1+(\delta _d^{LL})_{33}
\end{pmatrix}, \nonumber \\
M^2_{\tilde d_{RR}}&=m_{\tilde q}^2
\begin{pmatrix}
1+(\delta _d^{RR})_{11} & (\delta _d^{RR})_{12} & (\delta _d^{RR})_{13} \\
(\delta _d^{RR})_{12}^* & 1+(\delta _d^{RR})_{22} & (\delta _d^{RR})_{23} \\
(\delta _d^{RR})_{13}^* & (\delta _d^{RR})_{23}^* & 1+(\delta _d^{RR})_{33}
\end{pmatrix}, \nonumber \\
M^2_{\tilde d_{LR}}&=(M_{\tilde d_{RL}}^2)^\dagger =m_{\tilde q}^2
\begin{pmatrix}
(\delta _d^{LR})_{11} & (\delta _d^{LR})_{12} & (\delta _d^{LR})_{13} \\
(\delta _d^{LR})_{12}^* & (\delta _d^{LR})_{22} & (\delta _d^{LR})_{23} \\
(\delta _d^{LR})_{13}^* & (\delta _d^{LR})_{23}^* & (\delta _d^{LR})_{33}
\end{pmatrix},
\end{align}
where  $m_{\tilde q}$ is the average squark mass, and
$(\delta _d^{LL})_{ij}$, $(\delta _d^{LR})_{ij}$, $(\delta _d^{RL})_{ij}$, 
and $(\delta _d^{RR})_{ij}$ are  called as
the  mass insertion (MI) parameters.
The MI parameters are supposed to be much smaller than $1$.
%

The SUSY contribution by the gluino-squark box diagram to the dispersive part of the effective Hamiltonian for 
the $B_q$-$\bar B_q$ mixing  are written as \cite{Gabbiani:1996hi,Altmannshofer:2009ne} 
\begin{align}
M_{12}^{q,SUSY}&=A_1^q\Big [A_2\left \{ (\delta _d^{LL})_{ij}^2+
(\delta _d^{RR})_{ij}^2\right \} +
A_3^q(\delta _d^{LL})_{ij}(\delta _d^{RR})_{ij} \nonumber \\
&+A_4^q\left \{ (\delta _d^{LR})_{ij}^2+(\delta _d^{RL})_{ij}^2\right \} +A_5^q(\delta _d^{LR})_{ij}(\delta _d^{RL})_{ij}\Big ],
\label{bbbarmixing}
\end{align}
where 
\begin{align}
&A_1^q=-\frac{\alpha _S^2}{216m_{\tilde q}^2}\frac{2}{3}M_{B_q}f_{B_q}^2,
\qquad A_2=24xf_6(x)+66\tilde f_6(x),\nonumber \\
&A_3^q=\left \{ 384\left (\frac{M_{B_q}}{m_j+m_i}\right )^2+72 \right \} xf_6(x)+\left \{ -24\left (\frac{M_{B_q}}{m_j+m_i}\right )^2+36\right \} 
\tilde f_6(x), \nonumber \\
&A_4^q=\left \{ -132\left (\frac{M_{B_q}}{m_j+m_i}\right )^2 \right \} 
 xf_6(x),\quad A_5^q=\left \{ -144\left (\frac{M_{B_q}}{m_j+m_i}\right )^2-84\right \}\tilde f_6(x).
\label{Adef}
\end{align}
Here,  we use $x=m_{\tilde g}^2/m_{\tilde q}^2$, where
$m_{\tilde g}$ is the gluino mass.
For the cases of $q=d$ and  $q=s$, we take $(i, j)= (1,3)$ and
 $(i, j)= (2,3)$, respectively, where $m_1=m_d$, $m_2=m_s$ and $m_3=m_b$.
The loop functions
$f_6(x)$ and  $\tilde f_6(x)$ are given later in Eq.(\ref{loop}).

Let us discuss the  setup for the MI parameters in our analysis. 
For the case of $x\simeq 1$,  we estimate $A_2\simeq -1$,
$A_3^q\simeq 30$, $A_4^q\simeq -10$ and   $A_5^q\simeq 10$.
Therefore, 
 we consider the case that  $(\delta _d^{LL})_{ij}$ and
 $(\delta _d^{RR})_{ij}$ dominate $M_{12}^q$.
Actually, magnitudes of 
$(\delta _d^{LR})_{ij}$  and $(\delta_d^{RL})_{ij}$ are constrained
severely by the  $b\to s\gamma $ decay.


Including the double mass insertion, 
the transition amplitude of $b\to s\gamma$  
from the squark flavor mixing is given as  
\cite{Bertolini:1990if,Gabbiani:1996hi,Altmannshofer:2009ne} 
\begin{eqnarray}
A^{\rm SUSY}(b_L\to s_R\gamma )\propto
m_b M_3(x)(\delta _d^{LL})_{23}+m_{\tilde g}M_a(x)
(\delta _d^{LR})_{33}(\delta _d^{LL})_{23}
+m_{\tilde g}M_1(x)(\delta _d^{LR})_{23},
\label{bsgamma}
\end{eqnarray}
where 
functions 
$M_1(x)$, $M_3(x)$ and  $M_a(x)$  are given in Eq.(\ref{loop}).
At the electroweak scale, $(\delta _d^{LR})_{33}$ is given 
in terms of $\tan\beta$ and $\mu$ as 
\begin{equation}
(\delta _d^{LR})_{33}=
 m_b\frac{A_b-\mu \tan \beta}{m_{\tilde q}^2},
\label{mutan}
\end{equation}
where $A_b$ is the A-term given at the high energy scale.
In our numerical study, $A_b$ is taken to be $0$.
Since $m_{\tilde g}\gg m_b$,
   the magnitudes of $(\delta _d^{LR})_{23}$ and  $(\delta _d^{RL})_{23}$ 
 should be much smaller than
  $(\delta _d^{LL})_{23}$ and  $(\delta _d^{RR})_{23}$.


Therefore, we consider the contribution  from  $(\delta _d^{LL})_{ij}$ and
 $(\delta _d^{RR})_{ij}$ in  $M_{12}^q$.
In order to estimate the larger contribution of squark flavor mixing  
on $M_{12}^q$ with keeping smaller magnitudes of  MI parameters, 
we take  $|(\delta _d^{LL})_{ij}|=|(\delta _d^{RR})_{ij}|$.
This condition is derived  from  that 
 the coefficient $A_3^q$ is much larger than $A_2$.
On the other hand, we take    phases of these MI parameters 
$\theta _{ij}^{LL}$ and $\theta _{ij}^{RR}$  to be  different each other.
Therefore, we can parametrize the MI parameters  as follows:
\begin{equation}
(\delta _d^{LL})_{ij}=r_{ij}e^{2i\theta _{ij}^{LL}},
\qquad (\delta _d^{RR})_{ij}=r_{ij}e^{2i\theta _{ij}^{RR}}.
\label{MILLRR}
\end{equation}
Since magnitudes of $(\delta _d^{LR})_{23}$ and  $(\delta _d^{RL})_{23}$
are expected to be tiny from $b\to s\gamma$,  
we neglect them in our following calculations. 
%
%
Then, $r_{ij}$, $\theta _{ij}^{LL}$, and $\theta _{ij}^{RR}$ are related with
   the new physics contribution $h_q$ and $\sigma_q$.
Inserting Eq.(\ref{bbbarmixing}) with  Eq.(\ref{MILLRR}) into
 the following ratio 
\begin{equation}
\frac{M_{12}^{q,\text{SUSY}}}{M_{12}^{q,\text{SM}}}=h_qe^{2i\sigma _q}\ ,
\label{M12SMhsigma}
\end{equation}
we obtain two equations as follows: 
\begin{align}
&r_{ij}=\sqrt{\frac{h_q|M_{12}^{q,\text{SM}}|}{\left |A_1^q
\left (2A_2\cos 2\left (\theta _{ij}^{LL}-\theta _{ij}^{RR}\right )+
A_3^q\right )\right |}}~, \nonumber \\
&\theta _{ij}^{LL}+\theta _{ij}^{RR}=\sigma _q+\phi _q^\text{SM}+\frac{n\pi }{2},  \quad (n=0, \pm 1, \pm 2, \cdots ),
\label{r-theta-LL=RR}
\end{align}
where  $(\delta _d^{LR})_{ij}=(\delta _d^{RL})_{ij}=0$ is taken.
Here, we use the definition 
$2\phi_q^\text{SM}=\text{arg}(M_{12}^{q,\text{SM}})$ in the CKM basis.
The  numerical study of these  parameters are presented in the next section. 


There is another constraint for  MI parameters 
from the cEDM of the strange quark.
The T violation is expected to be observed 
 in the electric dipole moment of the neutron.
The experimental upper bound of the  electric dipole moment of the neutron
provides us the upper-bound of  the cEDM of the strange quark 
\cite{Hisano:2003iw,Hisano:2004tf,Hisano:2008hn}. 
The cEDM of the strange quark was discussed to  constrain the MI parameters 
 $(\delta _d^{LL})_{23}$ and $(\delta _d^{RR})_{23}$
 \cite{Endo:2010yt,Hisano:2003iw,Hisano:2004tf,Baker:2006ts}.

The cEDM of the strange quark is given by 
\begin{equation}
d_s^C=c\frac{\alpha _s}{4\pi }\frac{m_{\tilde g}}{m_{\tilde q}^2}\left (-\frac{1}{3}N_1(x)-3N_2(x)\right )
\text{Im}\left [(\delta _d^{LL})_{23}(\delta _d^{LR})_{33}(\delta _d^{RR})_{23}^*\right ],
\label{cedm}
\end{equation}
where $c$ is the QCD correction, and  $c=0.9$ is taken. 
The  $N_1(x)$ and $N_2(x)$ are given in Eq.(\ref{loop}).
By using 
Eq.(\ref{mutan}) and Eq.(\ref{MILLRR}) with $A_b=0$, 
 $d_s^C$ is rewritten  as
\begin{equation}
d_s^C=c\frac{\alpha _s}{4\pi }
\frac{m_{\tilde g}m_b\mu \tan \beta}{m_{\tilde q}^4}
\left (\frac{1}{3}N_1(x)+3N_2(x)\right )
r_{23}^2 
\sin 2(\theta_{23}^{LL}-\theta_{23}^{RR}).
\label{cedm2}
\end{equation}
Thus, the phase difference $(\theta_{23}^{LL}-\theta_{23}^{RR})$ 
is constrained 
from the experimental upper bound 
 $e|d_s^C|<1\times 10^{-25} \text{ecm}$
 \cite{Endo:2010yt,Hisano:2003iw,Hisano:2004tf,Baker:2006ts}.

The squark flavor mixing can be tested in
 the CP-violating  asymmetries  in the neutral $B$ meson decays. 
Since the $B_d^0\to J/\psi K_S$ process occurs at the tree level of SM, 
 the CP-violating  asymmetry  originates   from  $M_{12}^d$.
Although the $B_d^0\to \phi K_S$ and $B_d^0\to\eta 'K^0$ decays 
are penguin dominant ones,
their asymmetries also come from  $M_{12}^d$ in SM.
Then,  asymmetries of
 $B_d^0\to J/\psi K_S$,  $B_d^0\to \phi K_S$ and 
$B_d^0\to \eta 'K^0$ are expected to be same magnitude.
On the other hand, 
 if the squark flavor mixing  contributes to the decay 
 at the one-loop level, its magnitude could be  comparable 
to the SM penguin one
 in  $B_d^0\to \phi K_S$ and $B_d^0\to \eta 'K^0$, 
but it is tiny in $B_d^0\to J/\psi K_S$. 
Therefore, it is important to study carefully these asymmetries 
\cite{Endo:2004dc}.

Let us present the framework of these calculations.
The effective Hamiltonian for $\Delta B=1$ 
process is defined as 
\begin{equation}
H_{eff}=\frac{4G_F}{\sqrt{2}}\left [\sum _{q'=u,c}V_{q'b}V_{q's}^*\sum _{i=1,2}C_iO_i^{(q')}-V_{tb}V_{ts}^*
\sum _{i=3-6,7\gamma ,8G}\left (C_iO_i+\widetilde C_i\widetilde O_i\right )\right ],
\end{equation}
where the local operators are given as 
\begin{align}
&O_1^{(q')}=(\bar s_i\gamma _\mu P_Lq_j')(\bar q_j'\gamma ^\mu P_Lb_i),
\qquad O_2^{(q')}=(\bar s_i\gamma _\mu P_Lq_i')(\bar q_j'\gamma ^\mu P_Lb_j), \nonumber \\
&O_3=(\bar s_i\gamma _\mu P_Lb_i)\sum _q(\bar q_j\gamma ^\mu P_Lq_j),
\quad O_4=(\bar s_i\gamma _\mu P_Lb_j)\sum _q(\bar q_j\gamma ^\mu P_Lq_i), \nonumber \\
&O_5=(\bar s_i\gamma _\mu P_Lb_i)\sum _q(\bar q_j\gamma ^\mu P_Rq_j),
\quad O_6=(\bar s_i\gamma _\mu P_Lb_j)\sum _q(\bar q_j\gamma ^\mu P_Rq_i), \nonumber \\
&O_{7\gamma }=\frac{e}{16\pi ^2}m_b\bar s_i\sigma ^{\mu \nu }P_Rb_iF_{\mu \nu }, 
\qquad O_{8G}=\frac{g_s}{16\pi ^2}m_b\bar s_i\sigma ^{\mu \nu }P_RT_{ij}^ab_jG_{\mu \nu }^a,
\end{align}
where 
$P_R=(1+\gamma_5)/2$, $P_L=(1-\gamma_5)/2$, and $i$ and $j$ are color
 indices, and $q$ is taken to be $u,d,s,c$.
Here, $C_i$'s  $\widetilde C_i$'s are the Wilson coefficients, 
and $\widetilde C_i$'s
$\widetilde O_i$'s are  the operators by replacing  $L(R)$ with  $R(L)$ 
in $O_i$.
In our work,   $C_i$ includes both SM contribution and gluino   one,
such as  $C_i=C_i^{\rm SM}+C_i^{\tilde g}$, where
$C_i^{\text{SM}}$ is given in Ref.~\cite{Buchalla:1995vs} and
$C_i^{\tilde g}$ is  presented  as follows \cite{Endo:2004fx}:
\begin{align}
C_3^{\tilde g}&\simeq \frac{\sqrt{2}\alpha _s^2}{4G_FV_{tb}V_{ts}^*m_{\tilde q}^2}(\delta _d^{LL})_{23}
\left [-\frac{1}{9}B_1(x)-\frac{5}{9}B_2(x)-\frac{1}{18}P_1(x)-\frac{1}{2}P_2(x)\right ], \nonumber \\
C_4^{\tilde g}&\simeq \frac{\sqrt{2}\alpha _s^2}{4G_FV_{tb}V_{ts}^*m_{\tilde q}^2}(\delta _d^{LL})_{23}
\left [-\frac{7}{3}B_1(x)+\frac{1}{3}B_2(x)+\frac{1}{6}P_1(x)+\frac{3}{2}P_2(x)\right ], \nonumber \\
C_5^{\tilde g}&\simeq \frac{\sqrt{2}\alpha _s^2}{4G_FV_{tb}V_{ts}^*m_{\tilde q}^2}(\delta _d^{LL})_{23}
\left [\frac{10}{9}B_1(x)+\frac{1}{18}B_2(x)-\frac{1}{18}P_1(x)-\frac{1}{2}P_2(x)\right ], \nonumber \\
C_6^{\tilde g}&\simeq \frac{\sqrt{2}\alpha _s^2}{4G_FV_{tb}V_{ts}^*m_{\tilde q}^2}(\delta _d^{LL})_{23}
\left [-\frac{2}{3}B_1(x)+\frac{7}{6}B_2(x)+\frac{1}{6}P_1(x)+\frac{3}{2}P_2(x)\right ], \nonumber \\
C_{7\gamma }^{\tilde g}&\simeq -\frac{\sqrt{2}\alpha _s\pi }{6G_FV_{tb}V_{ts}^*m_{\tilde q}^2}
\Bigg [(\delta _d^{LL})_{23}\left (\frac{8}{3}M_3(x)-\mu \tan \beta \frac{m_{\tilde g}}{m_{\tilde q}^2}\frac{8}{3}M_a(x)\right )
+(\delta _d^{LR})_{23}\frac{m_{\tilde g}}{m_b}\frac{8}{3}M_1(x)\Bigg ], \nonumber \\
C_{8G}^{\tilde g}&\simeq -\frac{\sqrt{2}\alpha _s\pi }{2G_FV_{tb}V_{ts}^*m_{\tilde q}^2}
\Bigg [(\delta _d^{LL})_{23}\Bigg \{ \left (\frac{1}{3}M_3(x)+3M_4(x)\right ) \nonumber \\
&-\mu \tan \beta \frac{m_{\tilde g}}{m_{\tilde q}^2}\left (\frac{1}{3}M_a(x)+3M_b(x)\right )\Bigg \} 
+(\delta _d^{LR})_{23}\frac{m_{\tilde g}}{m_b}\left (\frac{1}{3}M_1(x)+3M_2(x)\right )\Bigg ].
\end{align}
The Wilson coefficients  $\widetilde C_i^{\tilde g}$'s are 
obtained by replacing $L(R)$ with $R(L)$ in  $C_i^{\tilde g}$'s.
The loop functions, which we use in our calculations, are summarized as 
\begin{eqnarray}
f_6(x)&=&\frac{6(1+3x)\log x+x^3-9x^2-9x+17}{6(x-1)^5}, \nonumber \\
\tilde f_6(x)&=&\frac{6x(1+x)\log x-x^3-9x^2+9x+1}{3(x-1)^5}, \nonumber\\
N_1(x)&=&\frac{3+44x-36x^2-12x^3+x^4+12x(2+3x)\log x}{6(1-x)^6}, \nonumber \\
N_2(x)&=&-\frac{10+9x-18x^2-x^3+3(1+6x+3x^2)\log x}{3(1-x)^6},\nonumber\\
B_1(x)&=&\frac{1+4x-5x^2+4x \log x+2x^2 \log x}{8(1-x)^4}, \nonumber \\
B_2(x)&=&x\frac{5-4x-x^2+2\log x+4x\log x}{2(1-x)^4}, \nonumber \\
P_1(x)&=&\frac{1-6x+18x^2-10x^3-3x^4+12x^3\log x}{18(x-1)^5}, \nonumber \\
P_2(x)&=&\frac{7-18x+9x^2+2x^3+3\log x-9x^2\log x}{9(x-1)^5}, \nonumber \\
M_1(x)&=&4B_1(x), \qquad \qquad    M_2(x)=-xB_2(x), \nonumber \\
M_3(x)&=&\frac{-1+9x+9x^2-17x^3+18x^2\log x+6x^3\log x}{12(x-1)^5}, 
\nonumber \\
M_4(x)&=&\frac{-1-9x+9x^2+x^3-6x\log x-6x^2\log x}{6(x-1)^5}, \nonumber \\
M_a(x)&=&\frac{1+9x-9x^2-x^3+\left (6x+6x^2\right )\log x}{2(x-1)^5},
 \nonumber \\
M_b(x)&=&-\frac{3-3x^2+\left (1+4x+x^2\right )\log x}{(x-1)^5}.
\label{loop}
\end{eqnarray}

The CP-violating  asymmetries $\mathcal{S}_f$ in Eq.~(\ref{sf}) are 
 calculated by using $\lambda_f$, which is given 
 for  $B_d^0\to \phi K_S$ and $B_d^0\to \eta 'K^0$ as follows:
\begin{align}
\lambda_{\phi K_S,\  \eta 'K^0}&=-e^{-i\phi _d}\frac{\displaystyle \sum _{i=3-6,7\gamma ,8G}
\left (C_i^\text{SM}\langle O_i \rangle+C_i^{\tilde g}\langle O_i \rangle+
\widetilde C_i^{\tilde g}\langle \widetilde O_i \rangle \right )}
{\displaystyle \sum _{i=3-6,7\gamma ,8G}
\left (C_i^{\text{SM}*}\langle O_i \rangle+C_i^{{\tilde g}*}
\langle O_i \rangle+\widetilde C_i^{{\tilde g}*}\langle\widetilde O_i \rangle \right )}~,  
\label{asymBd}
\end{align}
where  $\langle O_i \rangle$ is the abbreviation of
  $\langle f |O_i | B_q^0\rangle$.
It is noticed that  
$\langle\phi K_S|O_i|B_d^0\rangle=\langle\phi K_S|
\widetilde O_i|B_d^0\rangle $
and $\langle\eta' K^0|O_i|B_d^0\rangle=-\langle\eta' K^0|
\widetilde O_i|B_d^0\rangle$
because of  the parity of the final state.
We  have also $\lambda_{f}$ for  $B_s^0\to\phi \phi$ and $B_s^0\to\phi\eta '$
as follows:
\begin{align}
\lambda_{\phi \phi, \ \phi\eta '}&=e^{-i\phi _s}\frac{\displaystyle \sum _{i=3-6,7\gamma ,8G}
C_i^\text{SM}\langle O_i \rangle+C_i^{\tilde g}\langle O_i \rangle+
\widetilde C_i^{\tilde g}\langle \widetilde O_i \rangle}
{\displaystyle \sum _{i=3-6,7\gamma ,8G}
C_i^{\text{SM}*}\langle O_i \rangle+C_i^{{\tilde g}*}
\langle O_i \rangle+\widetilde C_i^{{\tilde g}*}\langle\widetilde O_i \rangle}~, 
\label{asymBs}
\end{align}
with 
$\langle\phi\phi|O_i|B_s^0\rangle=-\langle\phi\phi|\widetilde O_i|B_s^0\rangle $
and 
$\langle\phi\eta'|O_i|B_s^0\rangle=\langle\phi\eta'|\widetilde O_i|B_s^0\rangle$.

Although the $C_{8G}^{\tilde g}\langle O_{8G}\rangle$
dominates these decay amplitude, we take  account of 
other terms in our calculations.
Therefore, we estimate each  hadronic matrix elements
by using the factorization relations in Ref.~\cite{Harnik:2002vs}. 

We remark numerical input of phases $\phi_d$ and $\phi_s$.
The phase  $\phi_d$ is derived  from  the
 observed value  $\mathcal{S}_f=0.671\pm 0.023$ in $B_d^0\to J/\psi K_S$
\cite{PDG}
because  we have $\lambda_f=-e^{-i\phi_d}$ 
for $f=J/\psi K_S$.
On the other hand, we use the SM value of $\beta_s$ and
 the values of the new physics  parameters,  $h_s$ and $\sigma_s$ 
in Eq.(\ref{hsigmaLHCb})
 to estimate  $\phi_s=-2\beta_s+\text{arg}(1+h_s e^{2i\sigma_s})$. 
 We do not use the observed value of $\phi_s$ in $B_s^0\to J/\psi\phi$
 due to the large experimental error in Eq.(\ref{phis}).

 In our framework, we have taken the assumption 
$|(\delta _d^{LL})_{ij}|=|(\delta _d^{RR})_{ij}|$.
Let us compare our numerical results with the ones from another assumption,
 in which 
 $\delta _d^{RR}=0$ is taken.
Then, the MI parameters come from only left-handed soft scalar masses and
 phase is only one.
Now, the SUSY contribution by gluino-squark box diagram 
to the dispersive part of the effective Hamiltonian 
for the $B_q$-$\bar B_q$ mixing is simply written  as
\begin{equation}
M_{12}^{q,\text{SUSY}}=A_1^q A_2(\delta _d^{LL})_{ij}^2 .
\label{M12LL}
\end{equation}
Then,  the magnitude of the MI parameters and the phase are given as
\begin{align}
&r_{ij}=\sqrt{\frac{h_q|M_{12}^{q,\text{SM}}|}{\left |A_1^qA_2\right |}}~, \nonumber \\
&\theta _{ij}^{LL}=\frac{1}{2}\sigma _q+\frac{1}{2}\phi _q^\text{SM}+\frac{n\pi }{4}, \quad (n=0, \pm 1, \pm 2, \cdots) ,
\label{r-theta-RR=0}
\end{align}
instead of Eq.(\ref{r-theta-LL=RR}).
The numerical discussion are presented in the next section.

\section{Numerical analysis}
\label{sec:Numerical}

\begin{table}[h]
\begin{center}
\begin{tabular}{|c|c||c|c|}
\hline 
Input & & Input & \\
\hline 
$f_{B_s}$ & $(231 \pm 3\pm 15)~{\rm MeV}$ & $B_s(m_b)$ & $0.841 \pm 0.013 \pm 0.020$ \\
\hline 
$f_{B_s}/f_{B_d}$ & $1.209 \pm 0.007 \pm 0.023$ & $B_{s}/B_{d} $ & $1.01\pm 0.01 \pm 0.03$ \\
\hline 
$\hat \eta_B$ & $0.8393\pm 0.0034$ & $S_0(x_t)$ & $2.35$ \\
\hline 
$M_{B_s}$ & $5.3663\pm 0.0006$ GeV & $M_{B_d}$ & $5.27917\pm 0.00029$ GeV \\
\hline 
$m_d(m_b)$ & $(5.1\pm 1.3)\times 10^{-3}$ GeV & $m_s (m_b)$ & $0.085\pm 0.017$ GeV \\
\hline 
$m_b(m_b)$ & $4.248\pm 0.051$ GeV & $\tau _B$ & $(1.472_{-0.026}^{+0.024})\times 10^{-12}$ s \\
\hline
\end{tabular}
\caption{Parameters of the neutral $B$ meson mixing and quark masses
\cite{Lenz:2006hd}. }
\label{table}
\end{center}
\end{table}

Let us show numerical results.
The magnitude of the MI parameter $r_{23}$ is  calculated 
from Eq.~(\ref{r-theta-LL=RR}) or Eq.~(\ref{r-theta-RR=0}),
where $M_{12}^{s,\text{SM}}$ is fixed by putting relevant parameters
 shown in Table 1. 
The phases   $\theta _{23}^{LL}$ and $\theta _{23}^{RR}$
are   constrained  as seen in Eq.~(\ref{r-theta-LL=RR}) 
or Eq.~(\ref{r-theta-RR=0}). 
On the other hand, the cEDM of the strange quark constrains
 the  phase difference 
$\theta _{23}^{LL}-\theta _{23}^{RR}$ 
in the case of $|(\delta _d^{LL})_{23}|=|(\delta _d^{RR})_{23}|$
as seen in Eq.(\ref{cedm2}).
Especially, the constraint of the cEDM of the strange quark becomes severe 
 in the case of larger $\mu\tan\beta$.

In our following numerical calculations,
 we fix the squark mass and the gluino mass as
\begin{equation}
m_{\tilde q}=1000~\text{GeV},\qquad m_{\tilde g}=1000~\text{GeV}.
\end{equation}
The parameters of new physics, $h_s$ and $\sigma_s$ are given
in Eq.(\ref{hsigmaLHCb}).
Phase parameters $\theta _{23}^{LL}$ and $\theta _{23}^{RR}$
 are taken in the region $[0,\ \pi]$.
It is noticed that the squark mass $m_{\tilde q}$ is a variable
 for only Figure~1.
In Fig.~1(a), we show $r_{23}$ versus the  squark mass value  
 for  the case of $|(\delta _d^{LL})_{23}|=|(\delta _d^{RR})_{23}|$
with $\mu\tan\beta=5000$~GeV.
The region between the upper curve and lower one is excluded by
the constraint of phases $\theta _{23}^{LL}$ and $\theta _{23}^{RR}$
from the cEDM of the strange quark $d_s^C$.
The value of $r_{23}$ is  around $0.02$ at 
$m_{\tilde{q}}=1000~{\rm GeV}$.
Its value is almost same  for larger $\mu\tan\beta$ such as $20000$~GeV.


\begin{figure}[h!]
\begin{minipage}[]{1.0\linewidth}
\includegraphics[width=8cm]{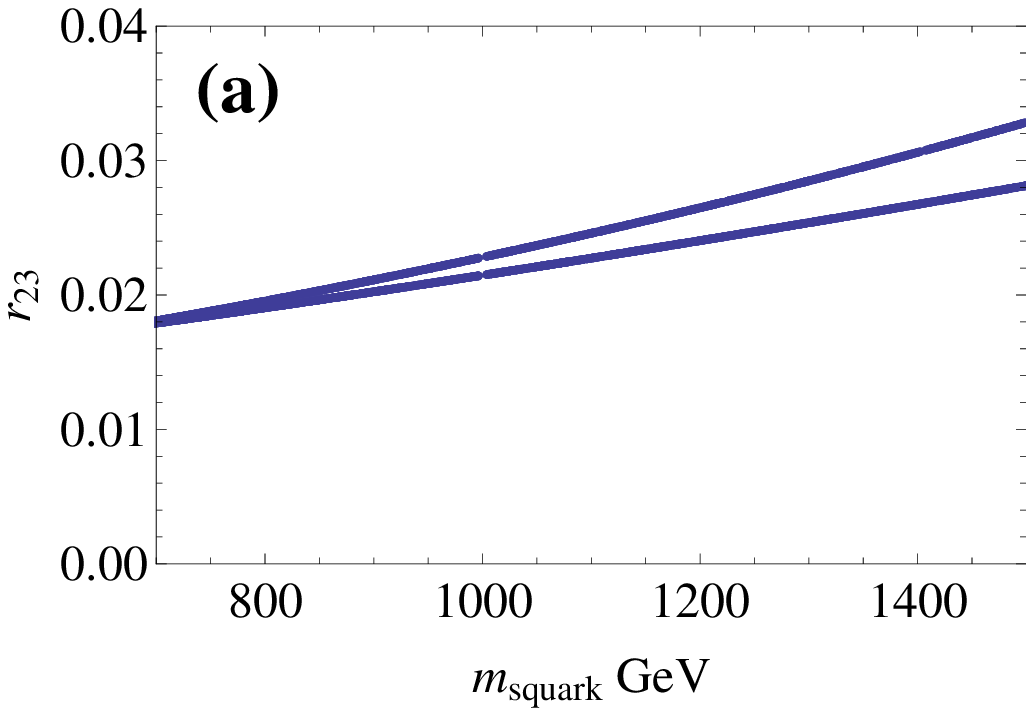}
\includegraphics[width=8cm]{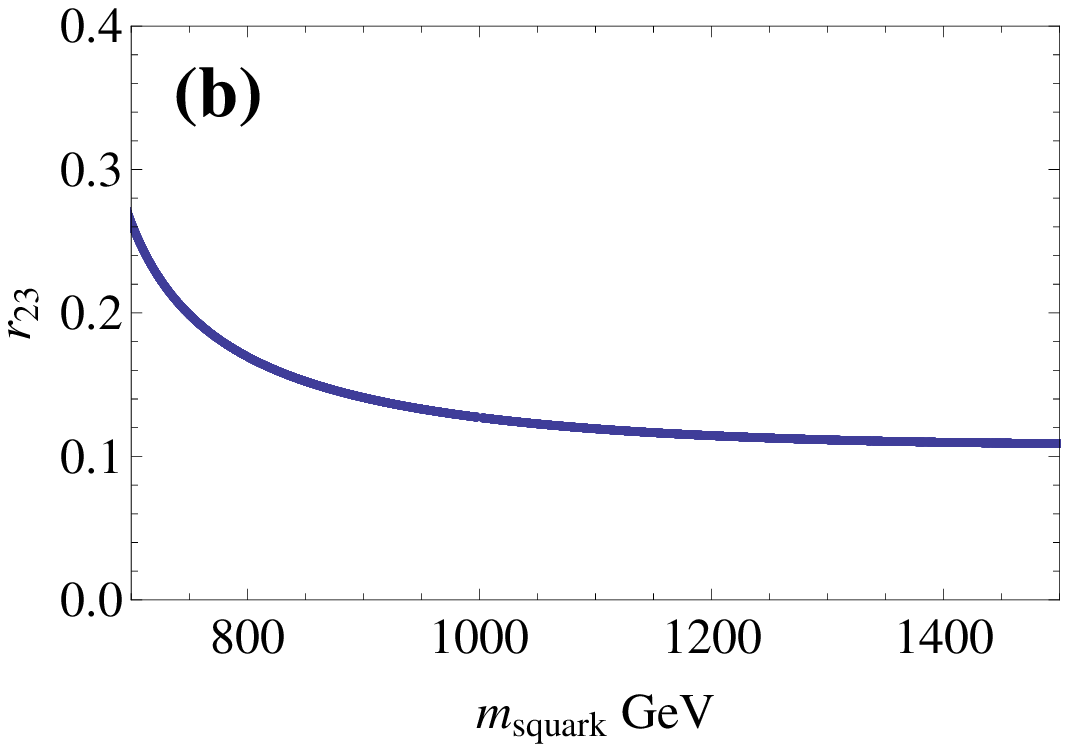}
\caption{The magnitude of $r_{23}$ versus squark mass 
at $\mu\tan\beta=5000~{\rm GeV}$
in the case of (a) $|(\delta _d^{LL})_{23}|=|(\delta _d^{RR})_{23}|$
 and (b) $(\delta _d^{RR})_{23}=0$.}
\label{mass-r}
\end{minipage}
\end{figure}

\begin{figure}[h!]
\begin{minipage}[]{1.0\linewidth}
\includegraphics[width=8cm]{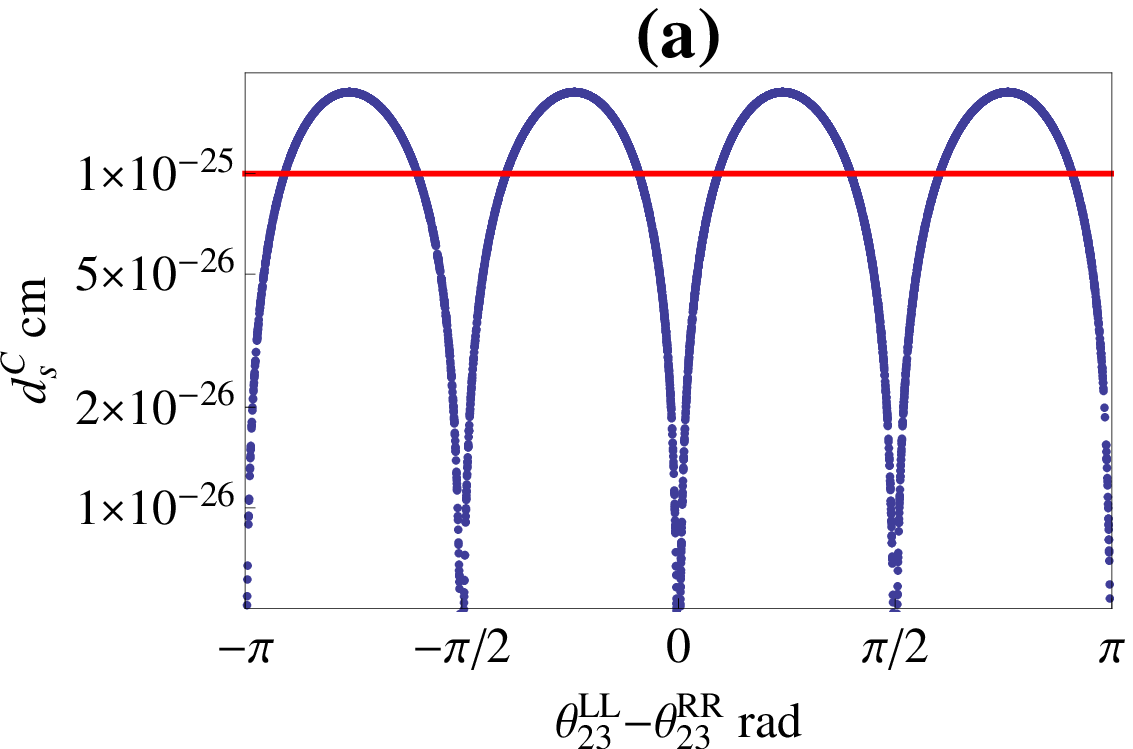}
\includegraphics[width=8cm]{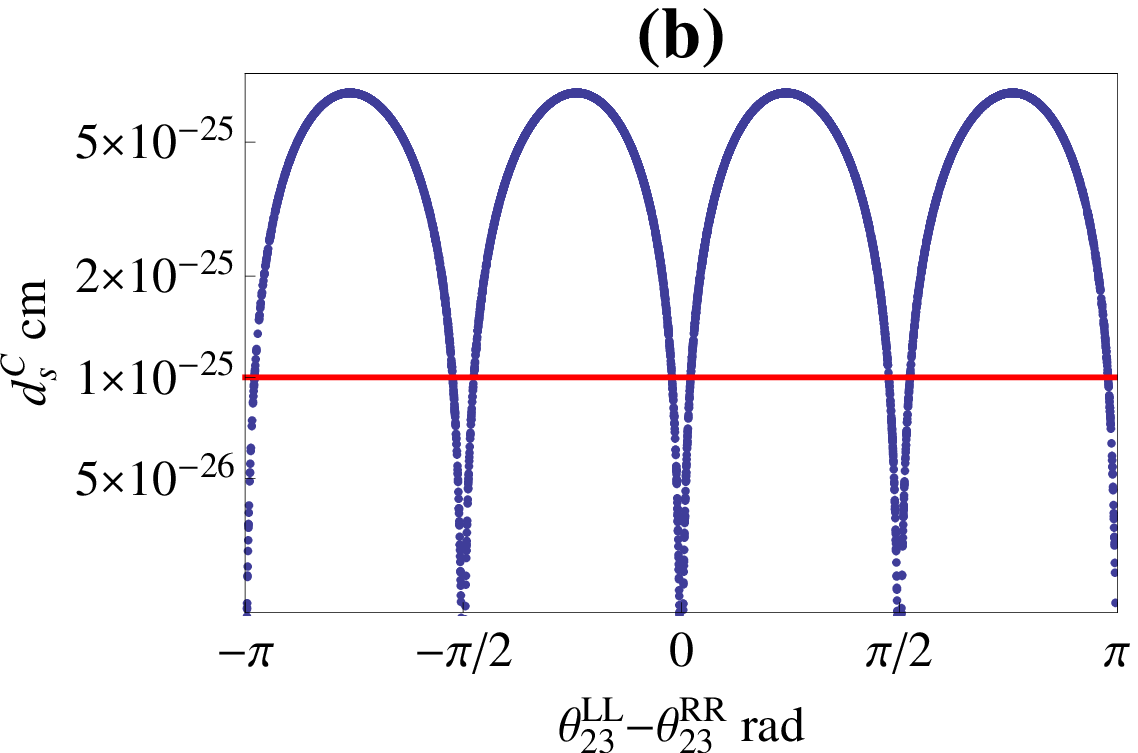}
\caption{The predicted  cEDM of the strange quark versus  the phase 
difference $\theta _{23}^{LL}-\theta _{23}^{RR}$ at 
(a) $\mu \tan \beta =5000$~GeV and (b) $\mu \tan \beta =20000$~GeV.
The experimental  upper bound is 
denoted by the red  horizontal line.}
\label{cEDM}
\end{minipage}
\end{figure}

In Fig.~1(b), we show   $r_{23}$
for  the case of $(\delta _d^{RR})_{23}=0$. 
There is no constraint from  $d_s^C$ because  of $(\delta _d^{RR})_{23}=0$. 
The value of $r_{23}$ is around  $ 0.13$
   at the $m_{\tilde q}=1000\ \text{GeV}$.
Thus, the obtained $r_{23}$ is six times larger 
 compared with the one for  $|(\delta _d^{LL})_{23}|=|(\delta _d^{RR})_{23}|$.

 The phases $\theta _{23}^{LL}$ and $\theta _{23}^{RR}$ 
 are constrained by the CP or T  violating experimental data.
 The cEDM of the strange quark in Eq.(\ref{cedm}) constrains
  the phase difference $\theta _{23}^{LL}-\theta _{23}^{RR}$.
Let us show the severe constraint from the cEDM of the strange quark.
In Figs.~2(a) and 2(b), the predicted values of $d_s^C$ are presented
 versus the phase difference 
$\theta _{23}^{LL}-\theta _{23}^{RR}$  at 
$\mu \tan \beta =5000$~GeV and 
$20000$~GeV, respectively, where 
the  red  horizontal line denotes  the experimental upper bound. 
It is noted that considerable tuning of the phase difference 
around $n\pi/2(n=0,\pm 1,  \cdots)$ is  required 
 for  $\mu \tan \beta=20000$~GeV.
These constraints affect the CP-violating  asymmetries
 in the non-leptonic $B$ meson decays.
On the other hand, for  the case of $(\delta _d^{RR})_{23}=0$, 
there is no  constraint from the cEDM of the strange quark.

\begin{figure}[h!]
\begin{minipage}[]{1.0\linewidth}
\includegraphics[width=8cm]{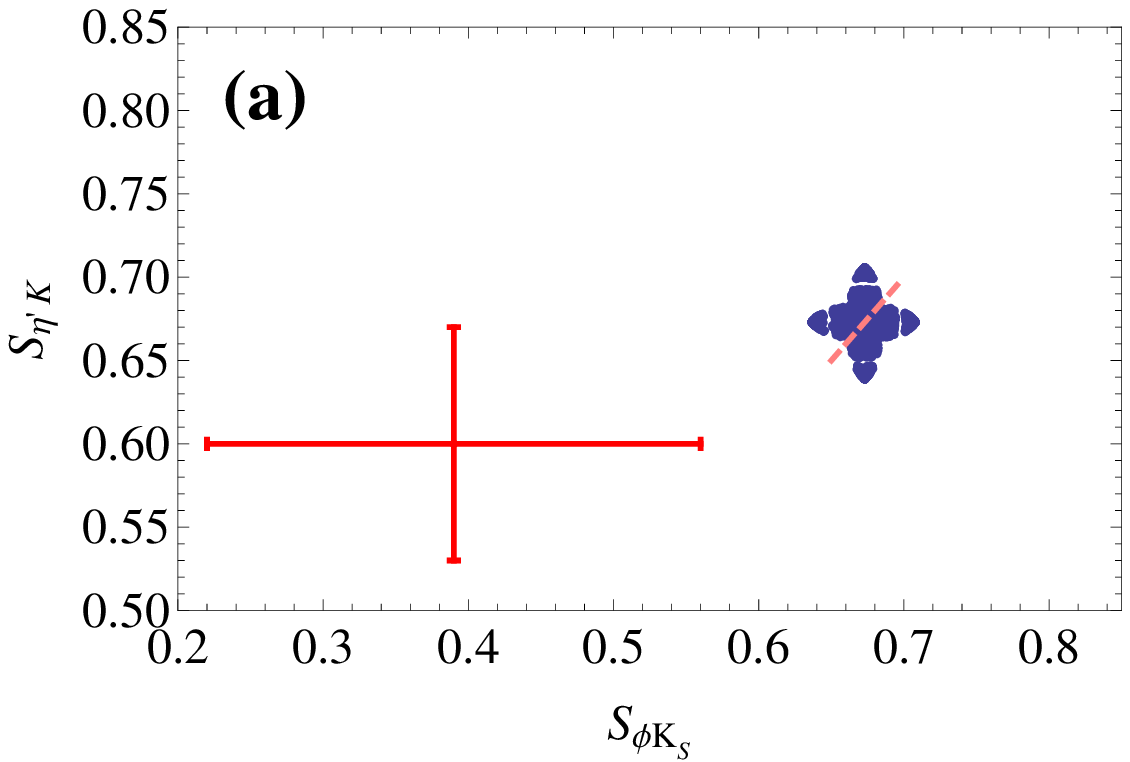}
\includegraphics[width=8cm]{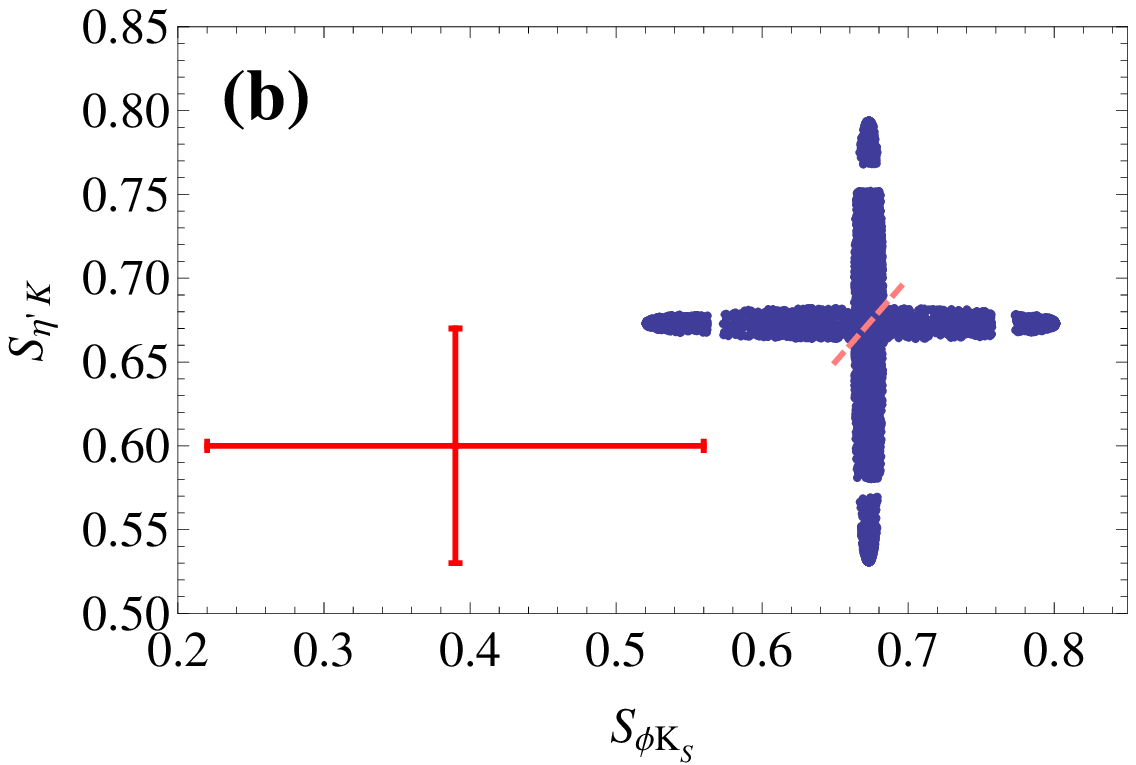}
\caption{Predicted  CP-violating  asymmetries of $B_d^0$ non-leptonic  decays
 in the case of 
$|(\delta _d^{LL})_{23}|=|(\delta _d^{RR})_{23}|$ at 
 (a)  $\mu \tan \beta =5000$~GeV and (b) $\mu \tan \beta =20000$~GeV.
The SM prediction $\mathcal{S}_{J/\psi K_S}=
\mathcal{S}_{\phi K_S}=\mathcal{S}_{\eta 'K}$ 
 is plotted by the  slant dashed lines. The 
experimental data with error bar is plotted by the red solid lines 
at $1~\sigma $ level.}
\label{LL=RR-SphiSeta}
\end{minipage}
\end{figure}

\begin{figure}[h!]
\begin{minipage}[]{1.0\linewidth}
\includegraphics[width=8cm]{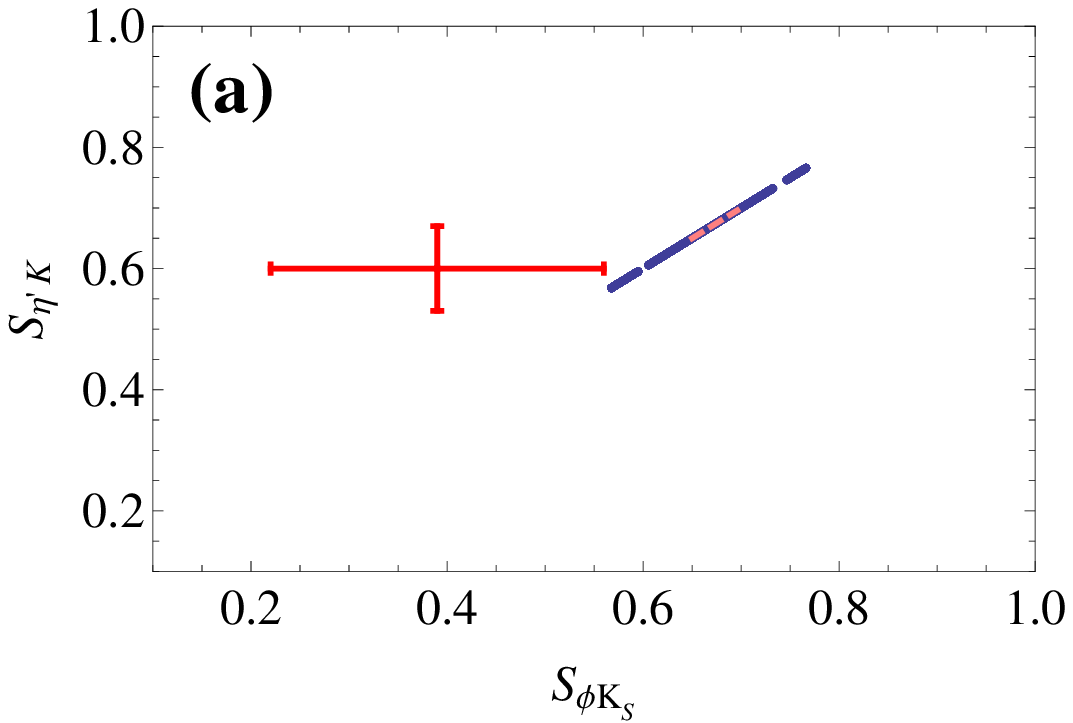}
\includegraphics[width=8cm]{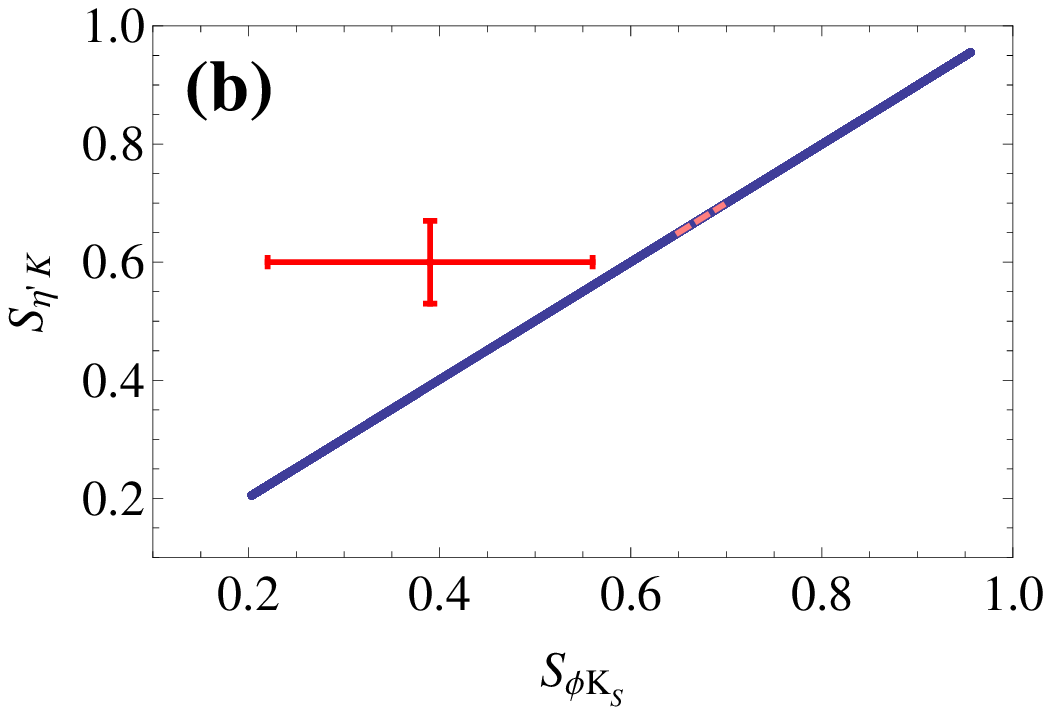}
\caption{Predicted  CP-violating  asymmetries   of $B_d^0$ non-leptonic  decays
 in the case of $(\delta _d^{RR})_{23}=0$ at 
 (a)  $\mu \tan \beta =5000$~GeV and (b) $\mu \tan \beta =20000$~GeV.
The SM prediction 
 denoted  by the  slant dashed line is  on the predicted line.}
\label{RR=0-SphiSeta}
\end{minipage}
\end{figure}

 By using the constrained MI parameters, we  predict
 the allowed region of the CP-violating asymmetries
 for the non-leptonic decays of the neutral $B$ mesons.
Let us discuss ${\mathcal S}_{f}$, which is the measure
 of the CP-violating asymmetry, 
for  $B_d^0\to {J/\psi  K_S}, \ {\phi K_S}, \  {\eta' K^0}$.
If there is no new physics, these   ${\mathcal S}_{f}$'s are predicted
 to be same ones.
On the other hand, 
 if the squark  flavor mixing contributes to the decay process 
 at the one-loop level, its magnitude is comparable to the SM penguin one
 in  $B_d^0\to \phi K_S$ and $B_d^0\to \eta 'K^0$,
but it is negligible small  in $B_d^0\to J/\psi K_S$.
Therefore, we expect  different   ${\mathcal S}_{f}$'s
 for  these  decays  from Eq.(\ref{asymBd}).

In Figs.~3(a) and 3(b),
 we show our predictions on the plane 
 $\mathcal{S}_{\phi K_S}$ and $\mathcal{S}_{\eta 'K^0}$
at $\mu \tan \beta =5000$~GeV and $\mu \tan \beta =20000$~GeV, respectively.  
The blue  regions denote  predicted ones
  from  our MI parameters $r_{23}$, $\theta_{23}^{LL}$,
and  $\theta_{23}^{RR}$,  which are constrained from $h_s$, $\sigma_s$
and $d_s^C$.
The red error bars of the horizontal and vertical solid lines 
 are  experimental values of $1~\sigma $ region 
in $\mathcal{S}_{\phi K_S}$-$\mathcal{S}_{\eta 'K}$. 
The slant dashed line denotes the SM prediction 
$\mathcal{S}_{J/\psi K_S}=\mathcal{S}_{\phi K_S}=\mathcal{S}_{\eta 'K}$,
where the observed value
 $\mathcal{S}_{J/\psi K_S}=0.671\pm 0.023$ is put.
 As seen  Fig.~3,
 the CP-violating  asymmetry is 
  deviated  a little from the SM prediction at  $\mu \tan \beta =5000$~GeV,
on the other hand, 
it can be significantly  deviated from the SM one
at $\mu \tan \beta =20000$~GeV.
Actually,
 it seems that 
the observed values  deviate from the SM predictions.
 We expect more  precise measurements of these  asymmetries
to find the new physics in the neutral $B$ meson decays.

Next, we discuss the case of $(\delta _d^{RR})_{23}=0$ 
in the decay $B_d^0\to \phi K_S$ and $B_d^0\to \eta 'K$. 
In Figs.~4(a) and 4(b), 
we show the predictions of the CP-violating  asymmetry
 on the  $\mathcal{S}_{\phi K_S}$-$\mathcal{S}_{\eta 'K}$ plane at
 $\mu \tan \beta =5000$~GeV and $20000$~GeV. 
In this case, there is no constraint from  the cEDM of the strange quark.
The allowed region
 is on the line, which is 
clearly different from the prediction in the case of
 $|(\delta _d^{LL})_{23}|=|(\delta _d^{RR})_{23}|$. 
 
\begin{figure}[h!]
\begin{minipage}[]{1.0\linewidth}
\includegraphics[width=8cm]{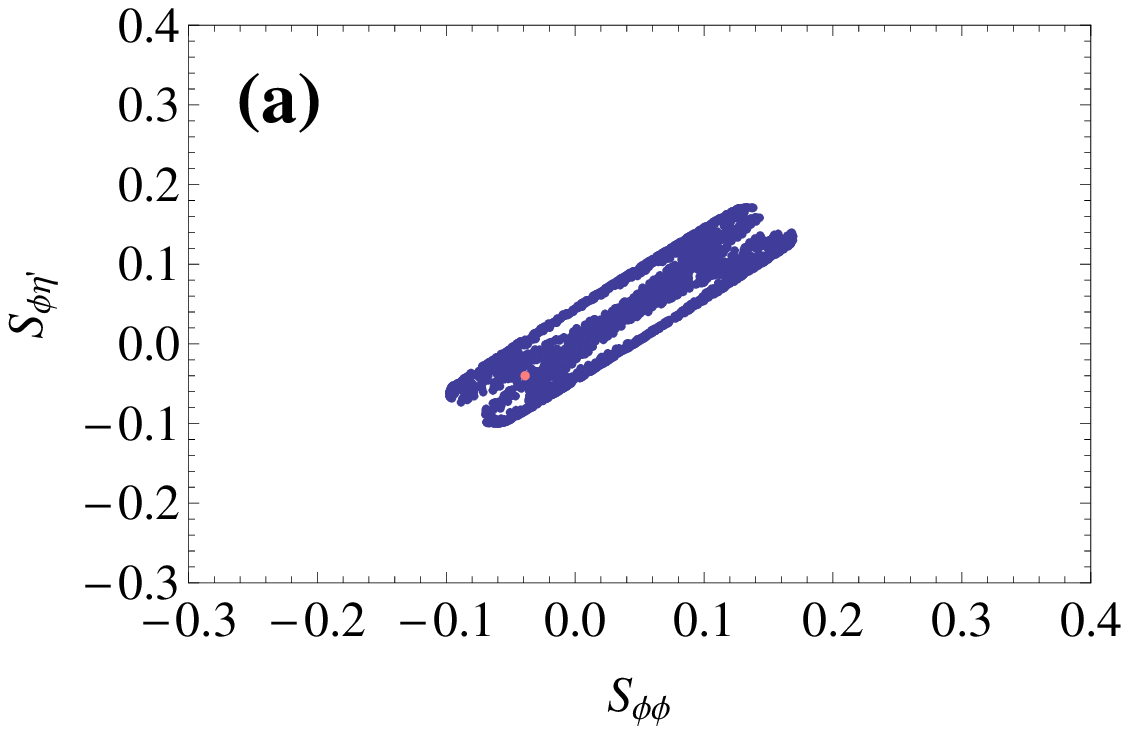}
\includegraphics[width=8cm]{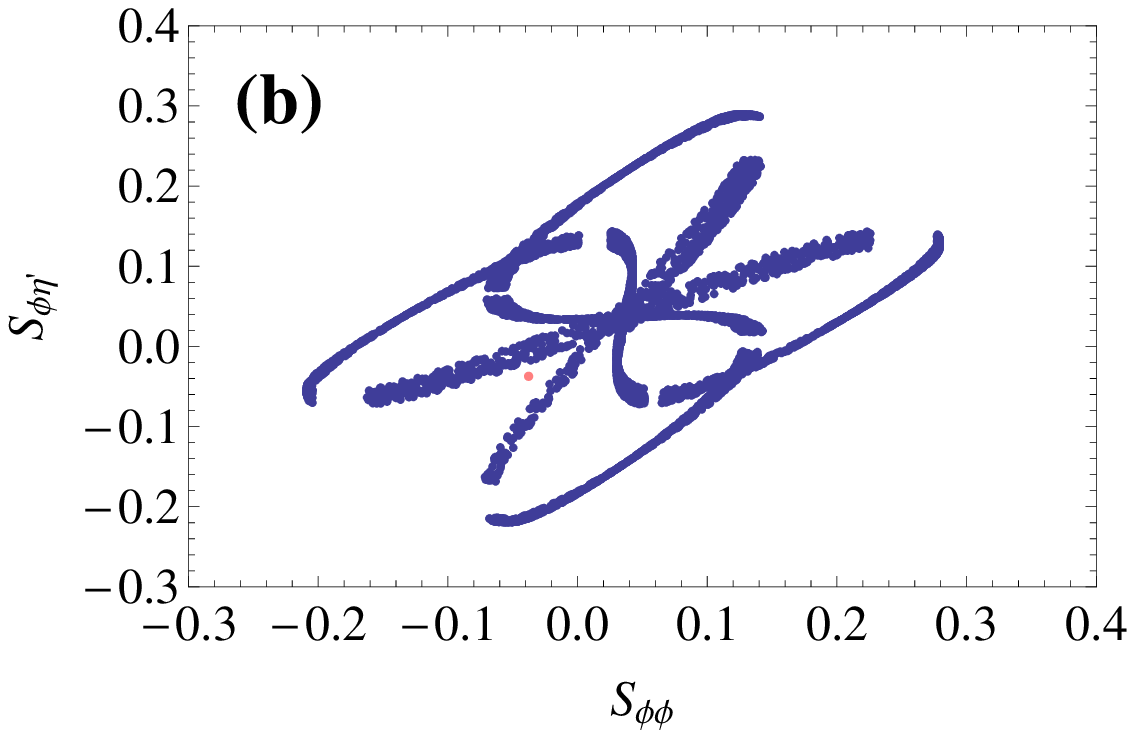}
\caption{Predicted  CP-violating  asymmetries
 of  $B_s^0$ non-leptonic decays  in the case of 
$|(\delta _d^{LL})_{23}|=|(\delta _d^{RR})_{23}|$
 at (a) $\mu \tan \beta =5000$~GeV and (b)  $\mu \tan \beta =20000$~GeV.
The central value of the SM prediction is 
plotted at $(-0.036, -0.036)$.}
\label{LL=RR-SphiphiSphieta}
\end{minipage}
\end{figure}

\begin{figure}[h!]
\begin{minipage}[]{1.0\linewidth}
\includegraphics[width=8cm]{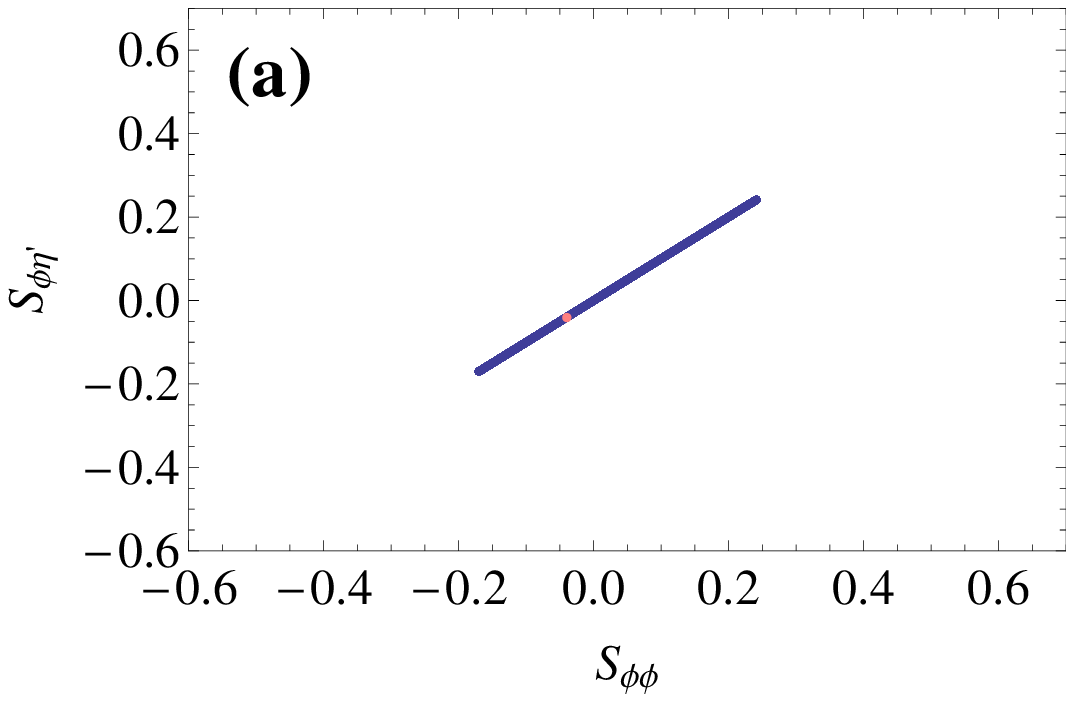}
\includegraphics[width=8cm]{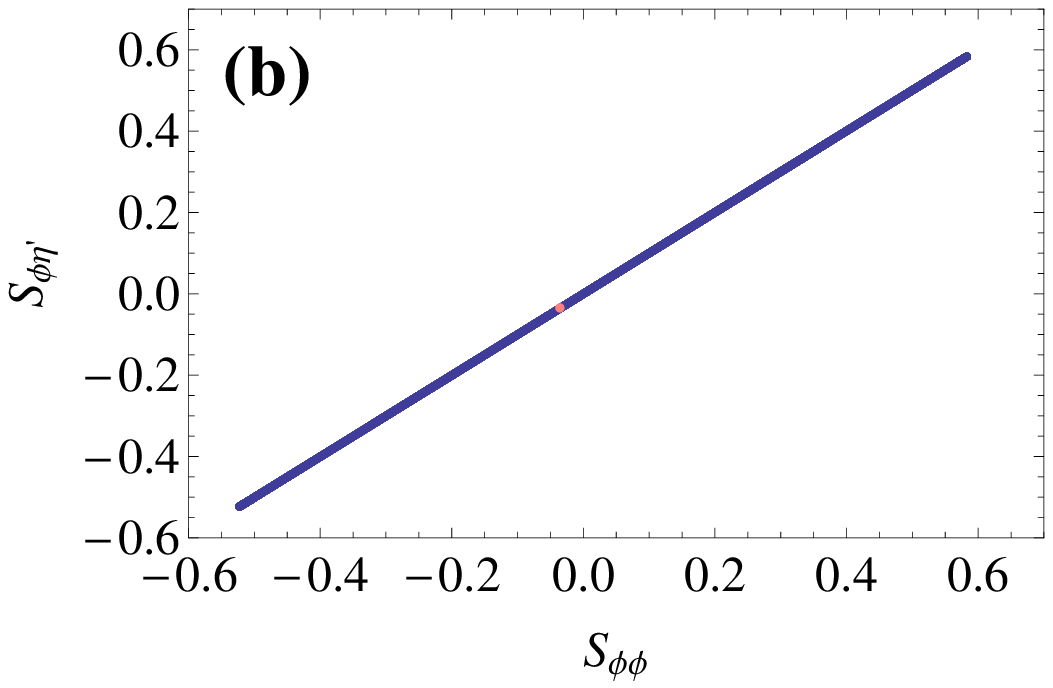}
\caption{Predicted  CP-violating asymmetries 
 of  $B_s^0$ non-leptonic decays  in the case of 
$(\delta _d^{RR})_{23}=0$
 at (a) $\mu \tan \beta =5000$~GeV and (b)  $\mu \tan \beta =20000$~GeV.
The central value of the SM prediction is 
plotted at $(-0.036, -0.036)$.}
\label{RR=0-SphiphiSphieta}
\end{minipage}
\end{figure}

 Since the LHCb observed the $B_s^0\to J/\psi\phi$ decay,
 we can now discuss the effect of the squark flavor mixing 
 on other CP-violating asymmetries such as the ones in
 $B_s^0\to \phi \phi $ and $B_s^0\to \phi \eta '$ decays.
In Figs.~5(a) and 5(b), 
we predict the CP-violating  asymmetries of
$\mathcal{S}_{\phi \phi }$ and $\mathcal{S}_{\phi \eta '}$ decays
at $\mu \tan \beta =5000$~GeV and $20000$~GeV,  respectively,
for the case of $|(\delta _d^{LL})_{23}|=|(\delta _d^{RR})_{23}|$. 
The blue  region denotes the predicted  region, and  
 the central value of the  SM prediction is plotted
 at  $(-0.036, -0.036)$, which is given in Eq.(\ref{SMBs}).
As seen in Fig.~5(b), the allowed region  on the 
 $\mathcal{S}_{\phi\phi}-\mathcal{S}_{\phi\eta'}$  plane is complicated
at $\mu \tan \beta =20000$~GeV
due to the severe phase constraint from  the cEDM of the strange quark
as seen in Fig.~2(a).

We also show  the result  of  the CP-violating  asymmetry 
for  the case of $(\delta _d^{RR})_{23}=0$. 
In Figs.~6(a) and 6(b), we predict the CP-violating  asymmetries  
at $\mu \tan \beta =5000$~GeV and $20000$~GeV.
  In this case, there is no constraint 
from  the cEDM of the strange quark.
These  asymmetries are expected to be observed at LHCb,
and then, new physics of squark flavor mixing  will be  testable. 


 
Finally, we discuss the constraint from the $b\to s\gamma$ decay, in which 
the transition amplitude  from the squark flavor mixing
is given in Eq.(\ref{bsgamma}).
The observed $b\to s\gamma$ branching ratio is 
$(3.60\pm  0.23)\times 10^{-4}$ \cite{PDG}, on the other hand 
the SM prediction is given as $(3.15\pm  0.23)\times 10^{-4}$
at ${\cal{O}}(\alpha_s^2)$
\cite{Buras:1998raa,Misiak:2006zs}.
Therefore, the contribution of our new physics should be suppressed
compared with the experimental data.
For  $|(\delta _d^{LL})_{23}|=|(\delta _d^{RR})_{23}|$
 with $(\delta _d^{LR})_{23}=(\delta _d^{RL})_{23}=0$,
we show the branching ratio including the contribution of the SM and 
the squark flavor mixing versus  $\mu \tan \beta$ in Figure 7(a),
where we neglect the error for the SM contribution.
Due to the phases 
 $\theta_{23}^{LL}$ and $\theta_{23}^{RR}$,
the predicted region is extended. 
As seen in Fig. 7(a), the contribution of the squark flavor mixing 
becomes seizable
 as   $|\mu \tan \beta|$ increases larger  than ${\cal O}(5000)$ GeV.
It is found that 
the contribution of the squark flavor mixing is consistent with
the experimental data  when we take account of the error for the
  SM prediction $(3.15\pm  0.23)\times 10^{-4}$.

For  the case of $(\delta _d^{RR})_{23}=0$,
 the contribution of the squark flavor mixing is larger than the one
 in the case  of $|(\delta _d^{LL})_{23}|=|(\delta _d^{RR})_{23}|$
as seen in  Figure 7(b). 
  The phase 
 $\theta_{23}^{LL}$  is somewhat constrained  to be consistent with
the experimental data for the large $|\mu \tan \beta|$.

In conclusion,  the $b\to s\gamma$ decay ratio
 hardly affects  our  predictions  of the   CP-violating asymmetries.

\begin{figure}[h!]
\begin{minipage}[]{1.0\linewidth}
\includegraphics[width=8cm]{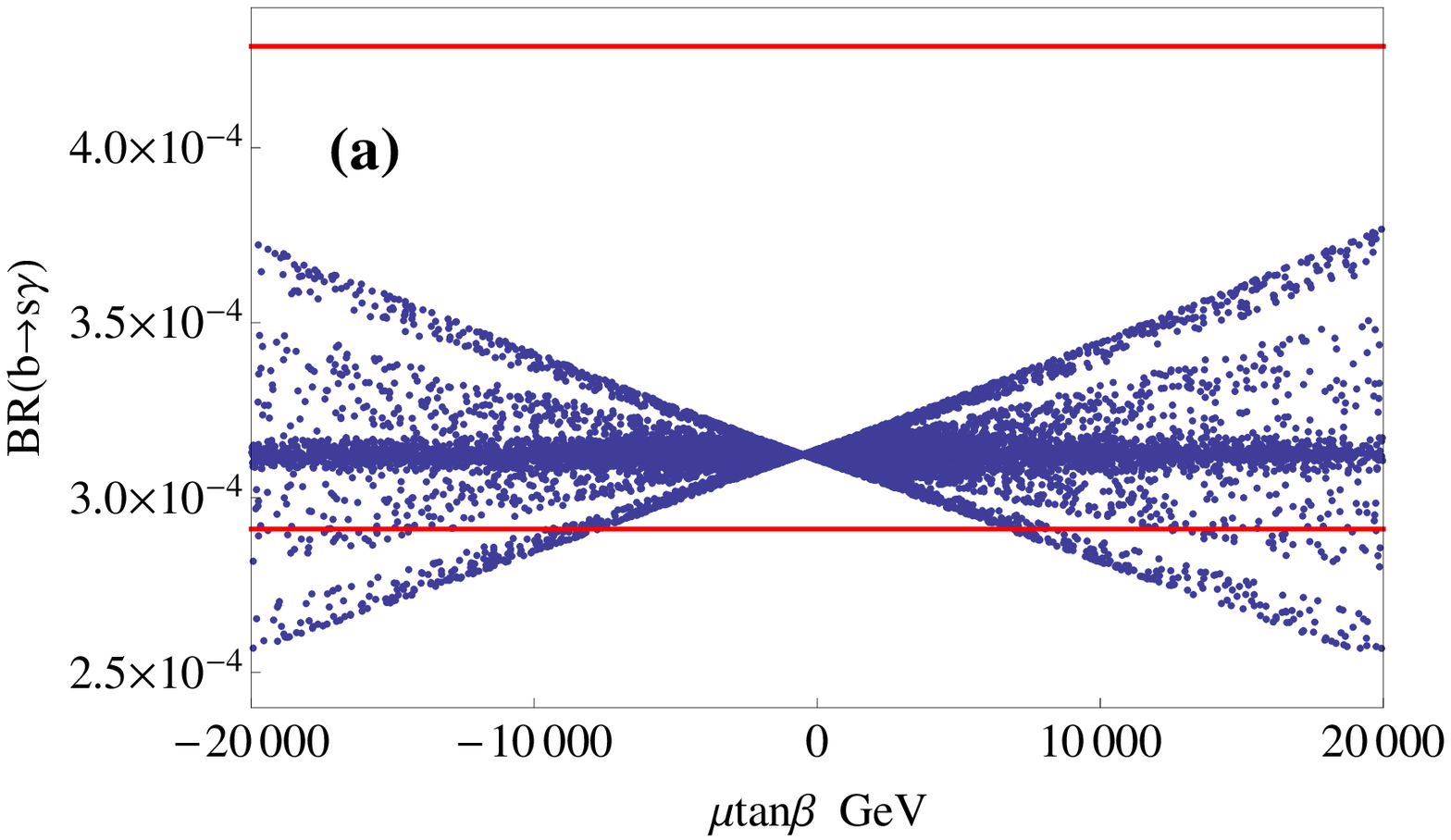}
\includegraphics[width=8cm]{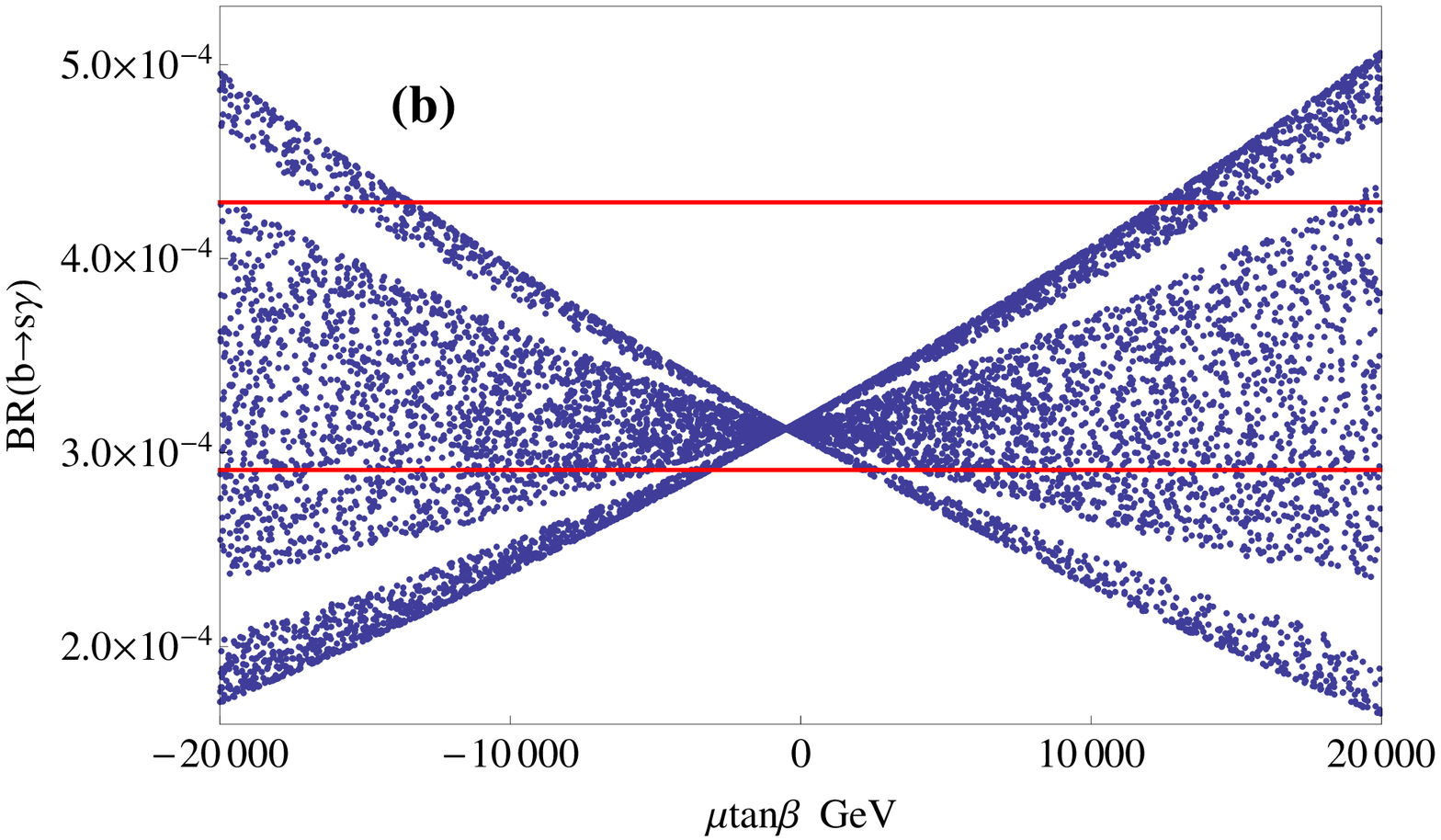}
\caption{
The $b\to s\gamma$ branching ratio
versus  $\mu \tan \beta$ for
(a) $|(\delta _d^{LL})_{23}|=|(\delta _d^{RR})_{23}|$
and (b) $(\delta _d^{RR})_{23}=0$.
The region between horizontal lines is allowed by
the experimental data at $3\sigma$.}
\end{minipage}
\end{figure}

\section{Summary and Discussion}
\label{sec:Summary}
We have discussed the contribution of the squark flavor mixing 
on the CP violation in the non-leptonic decays of $B_d^0$ and $B_s^0$ mesons
based on the recent LHCb data.
In our  predictions, 
we take account of the constraint from  the cEDM of the strange quark,
which is severe for  larger  $\mu \tan\beta$ such as $20000$~GeV.
CP-violating asymmetries of penguin dominated decays are the crucial test
 for  the squark flavor mixing.  We predict that
 the CP-violating  asymmetries $\mathcal{S}_f$ of $B_d^0\to \phi K_S$ and  
$B_d^0\to \eta 'K^0$ could deviate considerably 
from the one of  $B_d^0\to \phi K_S$ if $\mu\tan\beta\simeq 20000~{\rm GeV}$. 
Although these observed values seem to be different from the predictions of 
SM, more precise data are required in order to conclude
 the effect of the new physics.
Since  $B_s^0\to J/\psi \phi$ was observed at LHCb, 
we have also predicted  the asymmetries of 
 $B_s^0\to\phi \phi$ and $B_s^0\to\phi\eta '$.

 Since the global fit results of the CKMfitter do not guarantee 
the Tevatron anomaly,
 we should discuss  our input parameters of NP, $h_d$, $h_s$,
 $\sigma_d$ and $\sigma_s$ in Eq.(\ref{hdsigmaLHCb}) and Eq.(\ref{hsigmaLHCb})
in respect of 
  the like-sign dimuon charge asymmetry data at the D$\O$  Collaboration.
 Our parameters predict
 $A_{sl}^b=-(0.75 \sim 1.0)\times 10^{-3}$,
which is significantly deviated from  the SM prediction.
However, the experimental value of  the D$\O$  Collaboration
$-(7.87\pm 1.72\pm 0.93)\times 10^{-3}$
 still show  $3.5\sigma$ deviation from our predicted value.
 In conclusion, it is difficult to explain the Tevatron anomaly
  in our framework of the squark flavor mixing.


 The magnitudes of MI parameters may be important to build a flavor model
 such as the flavor symmetry.
 In our work, we obtained
$|(\delta _d^{LL})_{23}|=|(\delta _d^{RR})_{23}|\simeq 0.02$.
Putting the central values of CKMfitter,  
$(h_{d}\sim 0.3,\ \sigma_d \sim 1.8 \ {\rm rad})$, we obtain
  $|(\delta _d^{LL})_{13}|=|(\delta _d^{RR})_{13}|\simeq 0.008$.
The CP violation of the neutral $K$ meson also gives us 
 $|(\delta _d^{LL})_{12}|=|(\delta _d^{RR})_{12}|\leq  10^{-6}$.
Thus, we have the hierarchy of MI parameters
$|(\delta _d^{LL})_{23}|\geq |(\delta _d^{LL})_{13}|
\gg |(\delta _d^{LL})_{12}|$.
Such flavor structure of the squark mass matrix gives us a clue 
of the flavor symmetry.
We will discuss the flavor symmetry
 in the further coming paper.

\vspace{0.5 cm}
\noindent
{\bf Acknowledgement}

We thank M.~Endo, T.~Goto and R.~Kitano for useful discussions. 
Y.S. and M.T. are supported by JSPS Grand-in-Aid for Scientific Research,
22.3014 and 21340055, respectively.


\end{document}